\documentclass[runningheads]{llncs}
\usepackage{esvect}
\usepackage{float}
\usepackage[T1]{fontenc}
\usepackage{graphicx}
\usepackage{subcaption}
\usepackage{amssymb}
\usepackage{amsmath}
\usepackage{bm}
\usepackage{booktabs} 
\usepackage{multirow}

\begin{document}
\title{QEmbed: A Deep Learning Based Cardinality Estimator for Efficient Query Processing}
\titlerunning{QEmbed: Cardinality Estimator for Efficient Query Processing} 

\author{
Pooja Rajput\inst{1}\orcidID{0009-0009-1753-5056} \and
Suman Banerjee\inst{1}\orcidID{0000-0003-1761-5944}
}

\institute{
Indian Institute of Technology Jammu,
J \& K-181221, India\\
\email{\{2023rcs1016,suman.banerjee\}@iitjammu.ac.in}
}
 \maketitle              
\vspace{-1.5em}
\begin{abstract} Cardinality estimation is at the core of any commercial database system for efficient query processing. Over the decades, non-learning-based estimation techniques (e.g., histogram-based, sampling-based) have been widely used in both commercial and open-source database platforms. However, these techniques are only effective when the number of columns in a table is small, as they cannot properly capture dependencies between multiple attributes. Recently, learning-based approaches have been shown to perform significantly better than the heuristic methods that have been used for the past three decades. Despite this success, existing learned models often struggle to balance memory efficiency and accuracy when dealing with datasets that mix high and low cardinality attributes. In this paper, we propose a deep learning model formally called QEmbed. Our model is built upon the Masked Autoencoder for Distribution Estimation (MADE) auto-regressive framework to learn joint data distributions for selectivity estimation. To improve data representation and overcome the limitations of using a single encoding method, we design a hybrid encoding scheme that combines one-hot and embedding encodings. This hybrid design enables QEmbed to retain fine-grained attribute information for smaller domains while capturing compact semantic patterns for large, sparse domains.
We capture attribute correlations by factoring the joint data distribution into a series of conditional probabilities. This approach naturally accommodates both point and range queries. 
Through extensive experiments, we show that while QEmbed faces a latency trade-off on extremely wide schemas, it provides highly reliable cardinality estimates overall. A key advantage of our model is that it reduces extreme tail errors (maximum Q-errors), avoiding catastrophic estimation failures on complex, highly correlated workloads.

\end{abstract}
\section{Introduction} Cardinality estimation in database systems refers to estimating the output size of a relational algebra operation \cite{lohman2014query,leis2015good,lipton1990practical,olken1990random,wu2016sampling}. This is critical because accurate estimation directly leads to more efficient and relevant query execution plans. Over the past three to four decades, simple data-driven heuristics, such as histograms and sampling, have been widely used in commercial and open-source database systems \cite{ioannidis2003history}. However, literature has shown that these predictions are often suboptimal, sometimes resulting in estimation errors exceeding a factor of $10,000$ for complex queries.  

In the past decade, significant efforts have been made to build and employ machine learning and deep learning 
models for cardinality estimation. Supervised regression approaches, such as Multi-Set Convolutional Networks (MSCN) \cite{kipf2018learned} and Lightweight XGBoost (LW-XGB) \cite{zhu2025lightweight}, represent queries as feature vectors and train models to directly predict query selectivity \cite{kim2022learned}. These methods are effective when the training queries are representative of the workload but may struggle with previously unseen queries. Unsupervised approaches focus on modeling the joint probability distribution of relation attributes. Auto-regressive models like MADE, as used in NARU, capture correlations between attributes without relying on query logs and can handle equality and range predicates through techniques such as progressive sampling. 

DeepDB \cite{hasan2020deep} adopts a hybrid design that combines statistical techniques with neural networks, allowing it to handle high-dimensional datasets more efficiently. Nevertheless, representing the input space remains challenging because real-world datasets often contain a combination of low- and high-cardinality attributes, making it difficult to achieve both compact memory usage and accurate selectivity estimation.

 Earlier solutions such as NARU relied on deep autoregressive models to estimate the joint probability distribution of data attributes \cite{yang13deep}. This solution explored not only the unsupervised problem of density estimation using MADE models but also the supervised problem of selectivity estimation using a query-driven approach. In this study, binary and one-hot encoding schemes were also considered as ways to represent attribute values. 
 Both of these schemes can function well under certain conditions, but each has its shortcomings. In particular, one-hot encoding produces very sparse vectors and requires a heavy computational load when dealing with high-cardinality attributes. However, binary encoding uses space efficiently but lacks detail for low-cardinality attributes.
 
 To address these challenges, we present QEmbed with a hybrid encoding scheme used in MADE models.\\
 The main contributions of this work are summarized as follows:
\vspace{-0.3em}
\begin{itemize}
\item We investigate the cardinality estimation problem, which is fundamental to generating efficient query execution plans.

\item We propose QEmbed, a deep learning model that extends the MADE architecture with a hybrid encoding strategy to effectively represent datasets containing both low- and high-cardinality attributes.

\item We conduct extensive experiments on multiple real-world datasets and demonstrate that QEmbed achieves competitive cardinality estimation performance across diverse query workloads while substantially reducing extreme estimation errors compared with the Embed model.
\end{itemize}

\vspace{-0.3em}
\par The rest of the paper is organized as follows. Section \ref{Sec:Problem_Definition} formally defines the cardinality estimation problem. Section \ref{sec:model} presents the proposed model. Section \ref{Sec:Experiments} describes the experimental evaluation. Finally, Section \ref{Sec:CFD} concludes the paper and discusses future research directions.

\section{Problem Definition} \label{Sec:Problem_Definition}

Let $R$ be a relation with $m$ attributes $\{X_1, X_2, \dots, X_m\}$. Each attribute $X_i$ is associated with a finite domain $D_i$ derived from the distinct values present in the dataset. A selection query $Q$ is defined as a conjunction of $d$ predicates ($1 \le d \le m$) on these attributes:
\begin{equation}
    Q = \theta_1 \wedge \theta_2 \wedge \dots \wedge \theta_d
\end{equation}

Each predicate $\theta_i$ represents an equality condition ($X_i = a$), a range condition ($lb \le X_i \le ub$), or a membership condition ($X_i \in \{a_1, a_2, \dots, a_k\}$).

The cardinality of query $Q$, denoted as $\text{Card}(Q)$, is the number of tuples in $R$ that satisfy all predicates simultaneously:
\begin{equation}
    \text{Card}(Q) = |\{t \in R : Q(t) = 1\}|
\end{equation}

Correspondingly, the selectivity $\text{Sel}(Q)$ is the fraction of tuples in $R$ that satisfy the query:
\vspace{-0.2em}
\begin{equation}
    \text{Sel}(Q) = \frac{\text{Card}(Q)}{|R|}
\end{equation}

The main goal of selectivity estimation is to compute $\text{Sel}(Q)$ quickly without scanning the entire dataset. Traditional methods, such as histograms and sampling, usually rely on the attribute independence assumption. They approximate the joint probability by assuming that columns do not affect each other:
\vspace{-0.5em}
\begin{equation}
    P(X_1, \dots, X_m) = \prod_{i=1}^{m} P(X_i)
\end{equation}

Because this ignores real-world correlations between attributes, these methods often produce large errors for complex multi-attribute queries.
To address this limitation, we model the exact joint probability
distribution of the attributes. Using the MADE framework ~\cite{germain2015made}, we factorize the joint distribution into conditional probabilities:
\begin{equation}
    P(X_1, X_2, \dots, X_m) = \prod_{i=1}^{m} P(X_i \mid X_1, X_2\dots, X_{i-1})
\end{equation}

This allows the model to learn the actual dependencies between attributes directly. In this work, we focus on how different encoding schemes for categorical and numerical features operate within this model. We evaluate how these specific encoding choices impact the final estimation accuracy across different datasets.

\section{Proposed Methodology: QEmbed}
\label{sec:model}

\subsection{Overview of the QEmbed Architecture}
\begin{figure}[H]
    \centering
    \includegraphics[width=11cm,height=3.8cm]{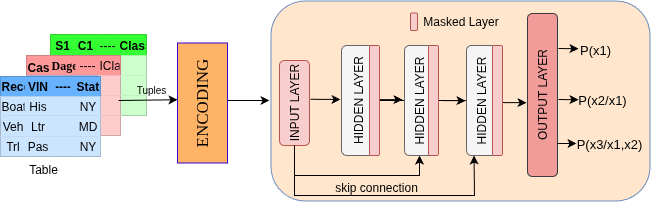}
    \caption{QEmbed: Architecture of the proposed model}
    \label{fig:qembed_architecture}
    \vspace{-1.5em}
\end{figure}

To achieve efficient and accurate query selectivity estimation, we introduce QEmbed, an auto-regressive deep learning network that treats cardinality estimation as an unsupervised density estimation problem. As shown in Fig.~\ref{fig:qembed_architecture}, the model consists of three main stages: Hybrid Encoding, Masked Processing, and Conditional Probability Estimation. Based on the MADE framework, QEmbed uses the probabilistic chain rule to factorize the joint probability of a relational tuple $\mathbf{x} = (x_1, x_2, \dots, x_m)$:
\vspace{-0.5em}
\begin{equation}
    P(\mathbf{x}) = \prod_{i=1}^{m} P(x_i \mid x_1, x_2, \dots, x_{i-1})
\label{eq:chain_rule}
\end{equation}
By conditioning each attribute only on its predecessors, this factorization allows the model to capture complex cross-attribute dependencies.
\vspace{-2em}

\subsection{Stage 1: Hybrid Encoding and Input Layer}
Relational databases contain mixed-type attributes. Encoding high-cardinality attributes with standard one-hot vectors creates massive, sparse vectors that are difficult to train. To solve this, QEmbed uses a dual-channel encoding approach controlled by a configurable cardinality threshold, $\tau$:

\begin{itemize}
    \item \textbf{Low-Cardinality Route ($|V_i| \le \tau$):} Domains with distinct values at or below the threshold use standard one-hot encoding. This preserves exact categorical boundaries for smaller domains.
    \item \textbf{High-Cardinality Route ($|V_i| > \tau$):} Domains exceeding the threshold bypass one-hot encoding. Instead, they pass through an embedding layer that maps them into a dense continuous vector $\mathbf{e}_i \in \mathbb{R}^d$.
\end{itemize}
Weights used during embedding initialization are drawn from a normal distribution. This hybrid model enables the system to learn semantic relationships
for larger sets and preserves a one-to-one mapping for smaller sets. Channels obtained through this are combined to create an input vector $h^{(0)}$.

\vspace{-1em}
\subsection{Stage 2: Masked Hidden Layers and Structural Connections}
\vspace{-0.5em}
The input vector $h^{(0)}$ is passed through a four-layer masked fully-connected neural network with 128 hidden neurons for each layer. 
This network size is chosen to capture all underlying data patterns while avoiding the high inference latencies characteristic of heavier models such as Transformers. The auto-regressive condition $P(x_i \mid x_{<i})$ is enforced using binary mask matrices $M^{(\ell)}$ on the weights. The formula describing feed-forward computation on a hidden layer $l$ is:

\begin{equation}
    \mathbf{h}^{(\ell)} = \text{ReLU}\left( \left(\mathbf{W}^{(\ell)} \odot \mathbf{M}^{(\ell)}\right) \mathbf{h}^{(\ell-1)} + \mathbf{b}^{(\ell)} \right)
\label{eq:masked_layer}
\end{equation}
Here, $W^{(\ell)}$ denotes the weight matrix, $b^{(\ell)}$ stands for the bias, and $\odot$ denotes the Hadamard product. Masking ensures that the network does not have the opportunity to use future attributes ($x_{\ge i}$). In addition, shortcuts from the beginning of the network to its end are used in order to make training faster. The input vector $h^{(0)}$ bypasses all layers and goes straight to the output layer without being multiplied by any masks.
Network weights are initialized using the Xavier uniform scheme, and biases are set to zero.
\vspace{-1em}
\subsection{Stage 3: Output and Unsupervised Training}
The final layer applies a softmax activation to output the estimated conditional probabilities, $\hat{P}(x_i \mid x_1, x_2,\dots, x_{i-1})$. We train the network in an unsupervised manner by minimizing the Negative Log-Likelihood (NLL) over the dataset $\mathcal{D}$:

\begin{equation}
    \mathcal{L}(\theta) = -\frac{1}{|\mathcal{D}|} \sum_{\mathbf{x} \in \mathcal{D}} \sum_{i=1}^{m} \log \hat{P}(x_i \mid x_1, , x_2, \dots, x_{i-1}; \theta)
\label{eq:nll_loss}
\end{equation}

Categorical attributes use cross-entropy loss, while binary attributes use binary cross-entropy. Because the input embeddings are updated jointly with the hidden layers during training, the model naturally groups highly correlated database values closer together in the latent space.
\vspace{-1em}
\subsection{Query Inference and Selectivity Estimation}
Once trained, QEmbed acts as an in-memory statistical summary of the database. For simple point queries, selectivity is calculated in $\mathcal{O}(1)$ time with a single forward pass by multiplying the output probabilities. For complex range queries where computing exact probabilities is too slow, QEmbed uses Progressive Sampling. This approach uses a Monte Carlo sampler to auto-regressively sample valid tuples from the learned distribution:

\begin{equation}
    \hat{Sel}(Q) \approx \frac{1}{N} \sum_{k=1}^{N} \prod_{i=1}^{m} \hat{P}(x_i^{(k)} \in Q_i \mid x_1^{(k)}, \dots, x_{i-1}^{(k)})
\label{eq:progressive_sampling}
\end{equation}

By relying entirely on the neural network's learned distribution, this method completely avoids physical table scans, providing fast and accurate cardinality estimates during query optimization.
\section{Experimental Evaluation} \label{Sec:Experiments}
In this section, we describe the experimental evaluation of the methodologies. 
We begin by describing the datasets used.
\vspace{-8pt}
\subsection{Datasets}
\vspace{-6pt}
In our experiments, we used the following five datasets:
\vspace{-4pt}
\begin{itemize}
\item \textbf{Forest}~\cite{forest}: A forest cover classification dataset with approximately 581,000 records and 55 categorical attributes, along with several numerical features. The correlated attributes and diverse feature space make it useful for evaluating cardinality estimation methods.

\item \textbf{Power}~\cite{bib21}: A household electricity consumption dataset with approximately 2.07 million records and 9 attributes. The continuous and highly correlated measurements provide a challenging workload for estimating cardinalities over numerical data.

\item \textbf{DMV}~\cite{dmv}: A real-world dataset containing approximately 11.6 million New York State vehicle, snowmobile, and boat registration records with 11 attributes. It includes diverse attribute domains and strong correlations, making it a challenging benchmark for cardinality estimation.

\item \textbf{Poker Hand}~\cite{poker}: This dataset contains 1,025,010 records with 11 attributes describing five-card poker hands. Strong dependencies among card attributes make it suitable for evaluating estimation accuracy on correlated data.

\item \textbf{Census}~\cite{census}: Based on the 1990 U.S. Census survey, this dataset contains approximately 2.45 million records and 61 demographic and socioeconomic attributes. Its high dimensionality and strong attribute correlations make it a challenging benchmark for cardinality estimation.
\end{itemize}

\vspace{-0.4cm}
\subsection{Baseline Methods}
In our experimentation, we compared our results with the following methods from the literature:
\begin{itemize}

\item \textbf{Sampling:} Sampling is a lightweight, widely used heuristic technique for cardinality estimation in query optimization. These techniques avoid scanning the entire dataset \cite{ioannidis1991sample} by running the query on a small sample of the data, and then scaling up the approximate cardinality estimation for the full dataset. Progressive sampling starts with a small sample and increases it gradually until the estimate becomes stable. This approach is particularly suitable for range queries \cite{wu2018random}.

\item \textbf{MaxHistDiff:} The MaxHistDiff approach generates multi-dimensional histograms to represent the dataset \cite{wang2020we}. We use the max-diff partitioning strategy, which identifies the largest difference between adjacent values and splits the data accordingly. During query estimation, MaxHistDiff identifies the relevant histogram buckets and aggregates their frequencies, assigning proportional weight to buckets that partially contribute \cite{poosala1997selectivity}.

\item \textbf{BayesNet:} BayesNet represents a probabilistic graphical model-based approach \cite{getoor2001selectivity}. The main aim of this model is to capture the dependencies among attributes for precise selectivity estimation. These models are used to find the joint distribution of the attributes and generate samples from these distributions to approximate query results \cite{yang13deep}.

\item \textbf{BayesCard:} This technique is an extended version of Bayesian networks to enhance cardinality estimation in databases. We follow the implementation from \cite{wu2020bayescard,tzoumas2011lightweight} which employs Bayesian networks with progressive sampling to find the cardinality of range queries. BayesCard allows us to obtain quick estimates more accurately, especially for complex queries.

\item \textbf{Transformer:} The Transformer is an auto-regressive model used to approximate the full joint distribution of the data without requiring independence assumptions. By using self-attention, it captures the correlations between attributes and supports both range and point queries \cite{zeng2024price}. This significantly improves the Q-error of the cardinality estimates \cite{tzoumas2011lightweight}.

\item \textbf{MADE:} MADE is a generative neural network that approximates the joint probability distribution of attributes using a masked auto-regressive architecture \cite{germain2015made}, where each prediction depends on previous attributes. For point queries, cardinalities are estimated by multiplying conditional probabilities and scaling by the table size; for range queries, the learned distribution guides progressive sampling \cite{yang13deep}.

\item \textbf{FACE:} FACE (Flow-based Auto-regressive Cardinality Estimator) uses normalizing flows to learn the joint probability distribution of relational data. Instead of grouping continuous or high-cardinality attributes into discrete buckets, it models them directly in a continuous space to capture complex data relationships. For range queries, FACE calculates exact probabilities using Cumulative Distribution Functions (CDFs) \cite{wang2021face}, avoiding the slow execution of Monte Carlo sampling while maintaining highly accurate estimates.
\end{itemize} 

\vspace{-0.7em}
\subsection{Details of the Experimentation}
\subsubsection{Experimental Setup and Hyperparameters:}
All our experiments were performed on a virtual machine (Ubuntu 20.04 LTS). This machine has an Intel Xeon processor (Cascade Lake with 40 CPU cores), 251 GB of RAM, and an NVIDIA Tesla V100 GPU with 32 GB of VRAM. CUDA version 11.8 was utilized, and PyTorch was used for building the machine learning model.

To ensure our results are fully reproducible, we set a fixed random seed of 0 for all operations. Because the objective of our cardinality estimator is to construct an accurate in-memory synopsis of a static database, the models were trained and evaluated on the complete datasets to capture the full underlying joint distributions. All models were trained for 50 epochs using a batch size of 256. We optimized the network using the Adam optimizer with an initial learning rate of $2\times10^{-4}$, which was dynamically scaled up to $1\times10^{-2}$ during batch processing to speed up model convergence. For the hybrid encoding layer, we set the embedding dimension to $d = 32$, initialized the embedding weights using a normal distribution with $\sigma = 0.02$, and set the default cardinality threshold to $\tau = 5$, though this threshold can be dynamically adjusted depending on the target database.
\vspace{-0.7em}

\subsubsection{Metrics for Evaluation:}
We apply the Q-error as the main criterion to evaluate estimation accuracy \cite{hasan2020deep,wang2020we}. The Q-error denotes the multiplicative factor between the estimate and the real value. To avoid division by zero, both the actual and estimated values are lower-bounded by 1. Formally, the Q-error for a particular query $Q$ can be calculated using the following formula:

\begin{equation}
    \text{Q-error} = \max \left( \frac{\widehat{Card}(Q)}{Card(Q)}, \frac{Card(Q)}{\widehat{Card}(Q)} \right)
\end{equation}

where $Card(Q)$ denotes the actual cardinality and $\widehat{Card}(Q)$ denotes the estimated one. The smaller the value of the Q-error, the more accurate the estimate is. We prefer using Q-error instead of other possible criteria, such as relative error, Mean Absolute Error (MAE), and Mean Squared Error (MSE). Q-error provides symmetrical treatment of under-estimation and over-estimation, does not depend on the size of the dataset, is immune to outliers, and is the standard metric used in the database literature that allows for easy comparison with existing methods. 
We report the median (50th percentile), 75th, 90th, and 95th percentile Q-errors, along with the mean and maximum Q-error, to evaluate estimation accuracy from both typical and worst-case perspectives. In addition, we report the inference latency of each method to assess its computational efficiency.

\vspace{-0.7em}
\subsubsection{Query Distribution:}
To evaluate performance, we generate a workload of random queries using a tuple-sampling method. For each query, the number of target attributes $f$ is chosen randomly, ranging from a minimum of 5 up to the total number of attributes $m$ present in the dataset. These $f$ distinct columns are selected without replacement. The filter values are extracted from a real tuple sampled uniformly at random from the dataset, ensuring that the queries reflect the actual data distribution. For attributes with a domain size of 10 or more, the filter operator is chosen uniformly from $\{\le, \ge, =\}$. For attributes with smaller domains, the operator is restricted exclusively to equality ($=$). Finally, any generated query that results in an actual cardinality of zero is discarded, ensuring the evaluation workload only contains queries with valid matching tuples.

\vspace{-0.7em}
\subsubsection{Training Model Performance}

\begin{figure*}[!ht]
    \centering
     \vspace{-1.5em}
    \begin{subfigure}{0.33\textwidth}
        \centering
        \includegraphics[width=\linewidth]{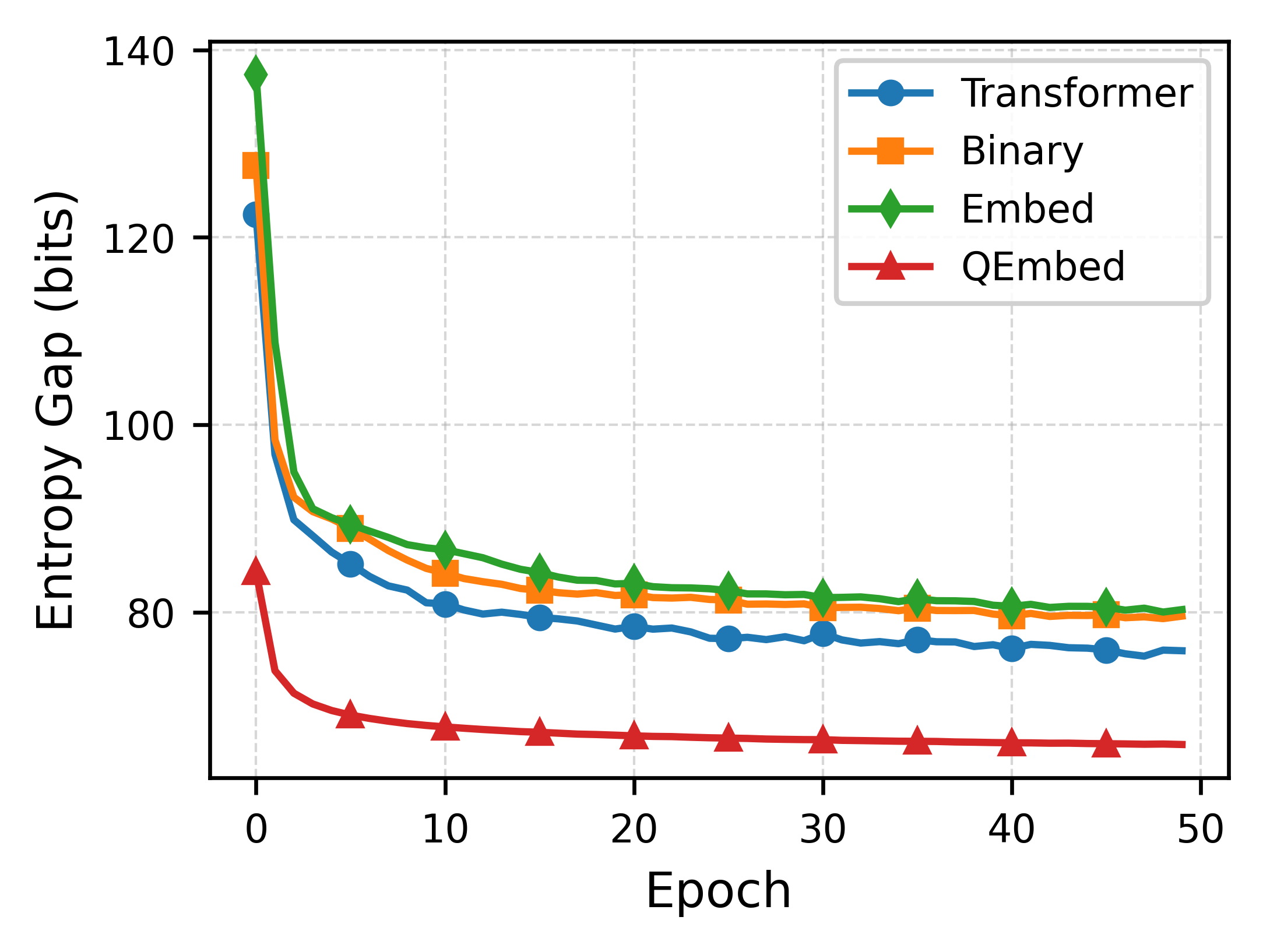}
        \caption{Forest dataset}
    \end{subfigure}
    \hfill
    \begin{subfigure}{0.33\textwidth}
        \centering
        \includegraphics[width=\linewidth]{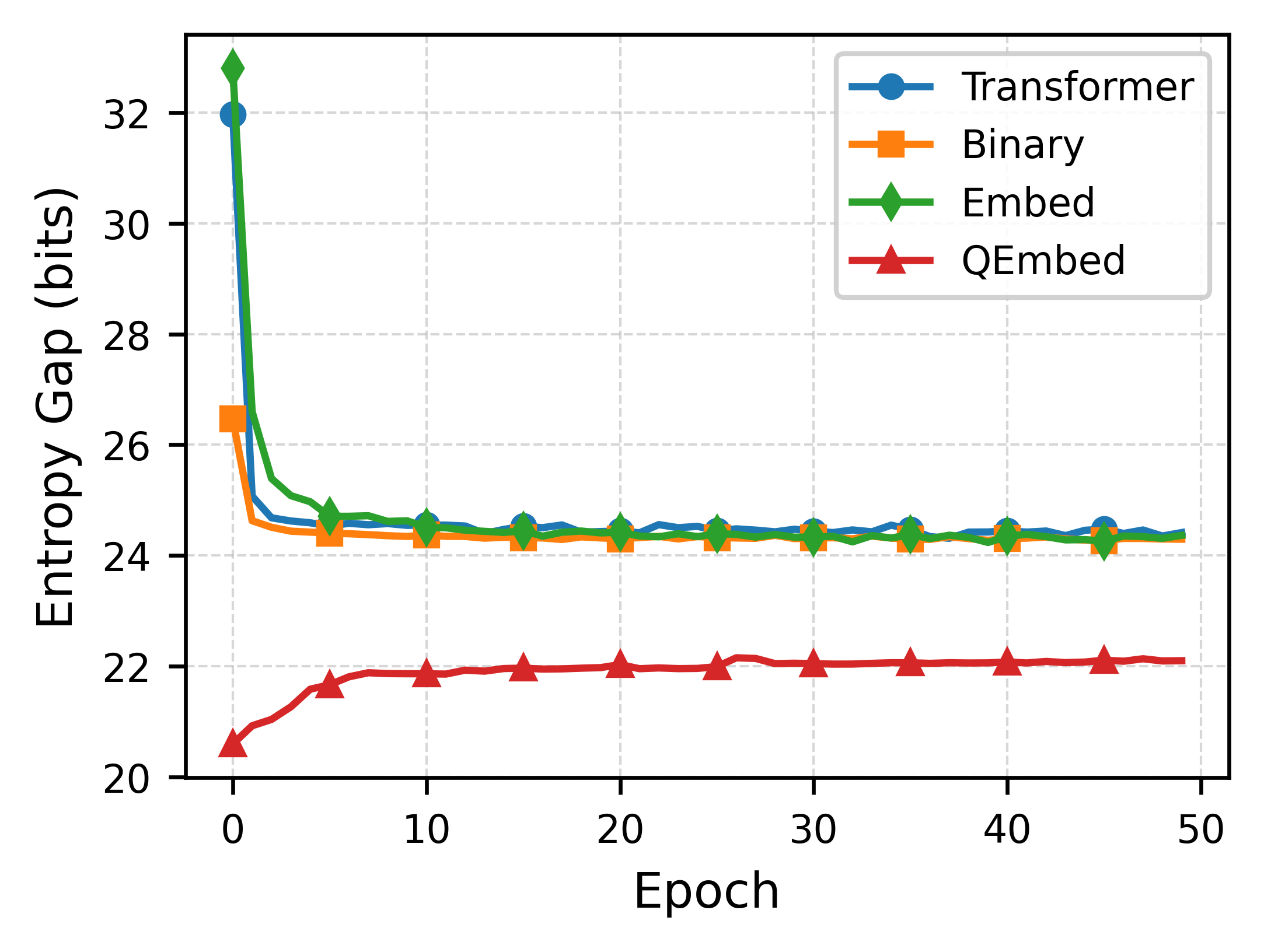}
        \caption{Power dataset}
    \end{subfigure}
    \hfill
        \begin{subfigure}{0.33\textwidth}
        \centering
        \includegraphics[width=\linewidth]{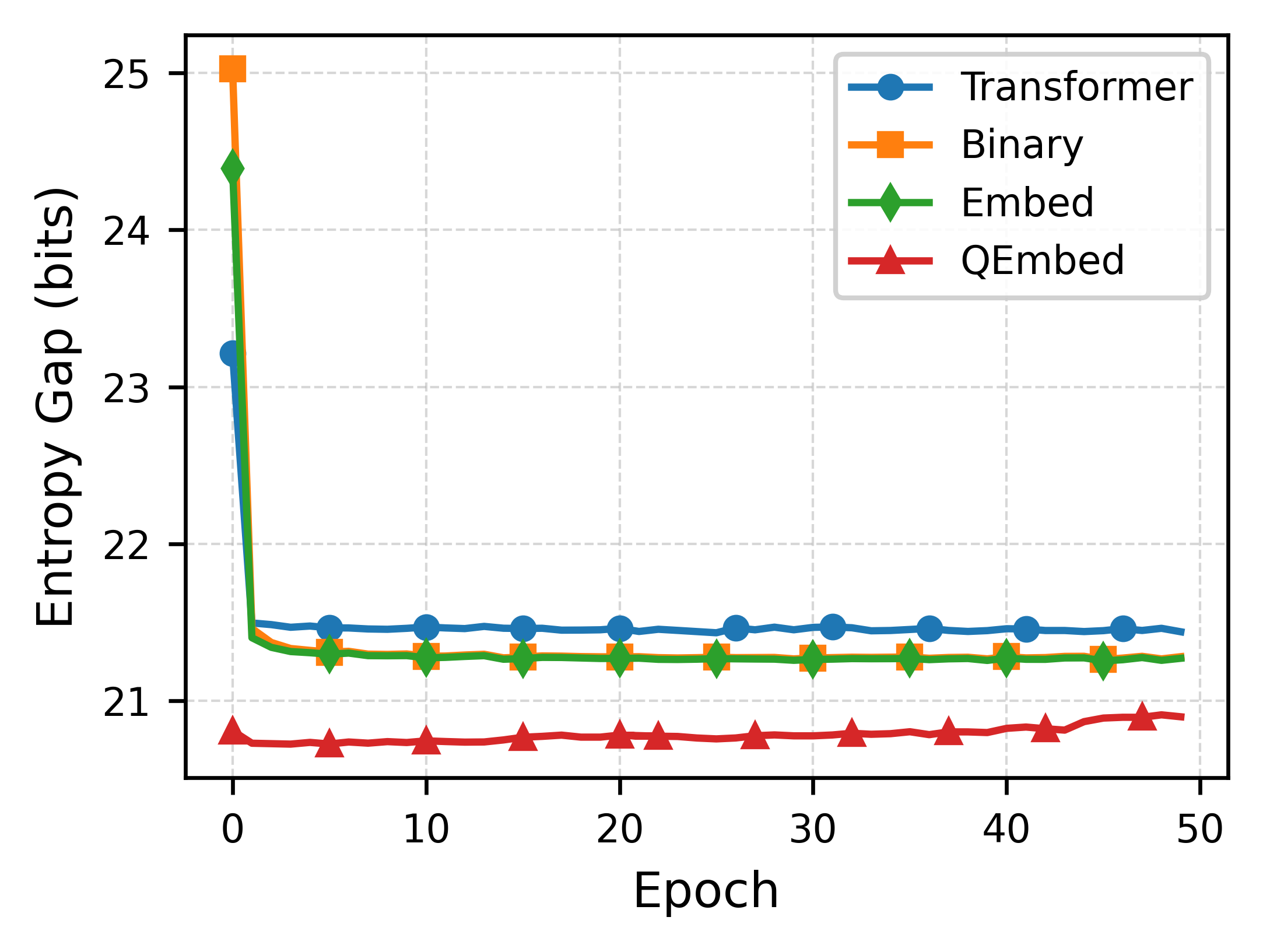} 
        \caption{DMV dataset}
    \end{subfigure}%
    \vspace{0.8em} 

    \hfill 
    \begin{subfigure}{0.33\textwidth}
        \centering
        \includegraphics[width=\linewidth]{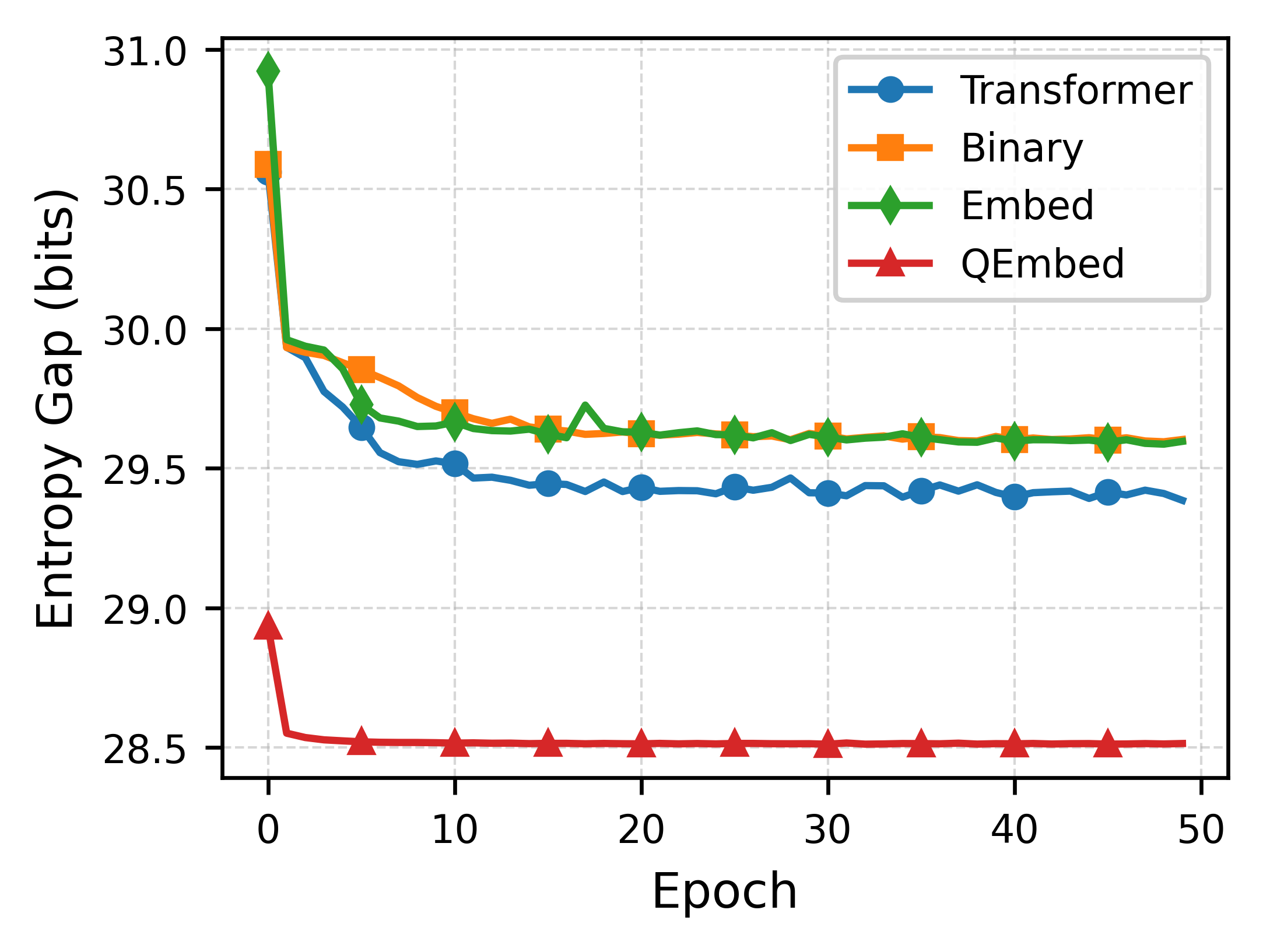}
        \caption{Poker dataset}
    \end{subfigure}
    \hfill
    \begin{subfigure}{0.33\textwidth}
        \centering
        \includegraphics[width=\linewidth]{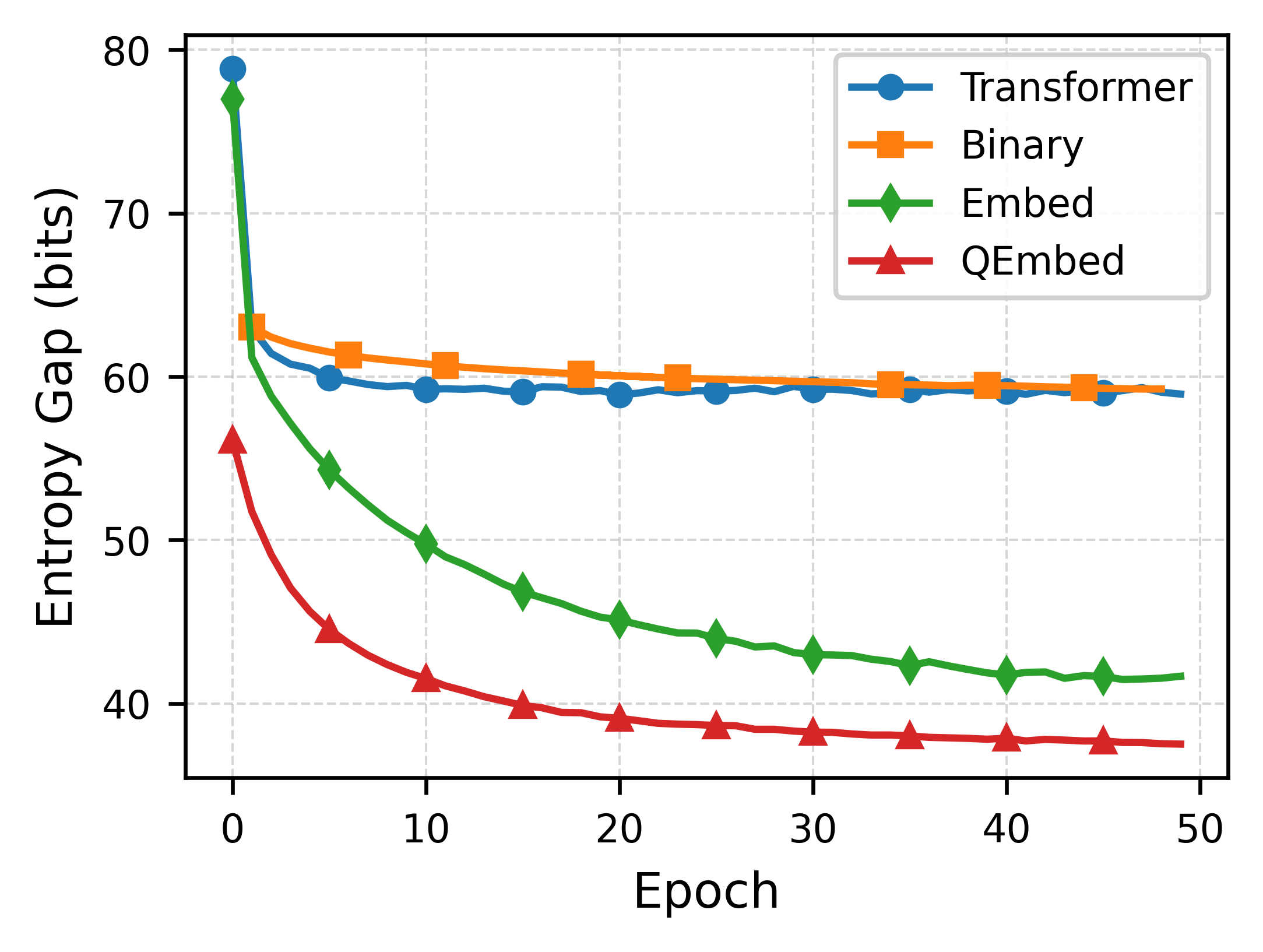}  
        \caption{Census dataset}
    \end{subfigure}
    \hfill \mbox{} 
    \caption{Comparison of training entropy gap (in bits) versus epochs for multiple models on the Forest, Power, DMV, Poker and Census datasets.}
    \label{fig:entropy_comparison}
    \vspace{-2em}
\end{figure*}

Figure~\ref{fig:entropy_comparison} compares the training entropy of the Transformer, Binary, Embed, and QEmbed models on five datasets. Across all datasets, QEmbed consistently achieved the lowest training entropy and reached a low-entropy solution earlier than the other models.
For the Forest dataset (Figure~\ref{fig:entropy_comparison}a), QEmbed reached a final entropy of about 75.29 bits at epoch 49. It finished training in 22,321 seconds, whereas the Binary and Embed models took much longer, requiring 43,120 and 44,710 seconds, respectively. On the Power dataset (Figure~\ref{fig:entropy_comparison}b), QEmbed maintained the lowest training entropy from start to finish. Although a slight increase in entropy was observed after the initial epochs, it consistently remained lower than that of the competing models, indicating stable optimization.

QEmbed trained on DMV in 22,804 seconds and achieved the lowest entropy of any model tested (19.24 bits). Binary and Embed needed over 449,767 and 481,311 seconds respectively to reach only 20.24 and 20.22 bits — roughly 20× slower for a slightly worse result.
Looking at the Poker dataset (Figure~\ref{fig:entropy_comparison}d), QEmbed reached a final entropy of 28.51 bits and completed training in just 1,268 seconds. 
Finally, for the Census dataset (Figure~\ref{fig:entropy_comparison}e), QEmbed again kept the lowest training entropy and showed stable progress during training.

This improved performance comes from our proposed hybrid encoding, which combines dense embeddings with binary encoding. This approach helps the model reach a low-entropy solution more efficiently while maintaining the lowest training entropy across all datasets. Overall, these results show that QEmbed offers a faster, more efficient training method for neural cardinality estimation across all five datasets.
\vspace{-1em}
\subsubsection{Effects of Encoding:}
In our experiments, we evaluated the impact of different encoding schemes (One-hot, Binary and Embed) on the model's performance. The results highlight distinct trade-offs:

\begin{itemize}
\vspace{-0.5em}
    \item \textbf{Space:} In the one-hot encoding scheme, the storage requirement is extremely high because each category requires a separate vector dimension, making it infeasible to encode high-cardinality features.      
    Binary encoding uses binary numbers to represent categories; therefore, the dimension increases logarithmically with an increase in the domain space. The embedding technique encodes categories in dense vectors of fixed dimension $d$.
  
    \item \textbf{Training Time:} The one hot encoding scheme is known to have very high training times because of the high dimensions of the input space, prolonging training due to the extensive memory overhead and large parameter space.
   However, Embedding is the most time-efficient method for large and skewed categorical domains.
    \item \textbf{Query Performance:} One-hot encoding works well for small domains due to the explicit representation of each category. However, as domain size increases, sparsity increases, which degrades generalization and slows down inference time. Binary encoding reduces this sparsity, making query execution more efficient for high-cardinality data. Embedding techniques excel by capturing the latent semantic similarity between attributes, which significantly aids in modeling complex data dependencies.
    \item \textbf{Data Type Suitability:} One-hot encoding is best suited for low-cardinality categorical features. Binary encoding is appropriate for medium-to-high cardinality features or skewed data. Embedding is the most flexible approach, efficiently handling both high-cardinality and heavily skewed datasets.
\end{itemize}
\vspace{-0.55cm}
\subsection{Experimental Results and Discussion}
We evaluate the performance of our proposed QEmbed model across five distinct datasets: Forest, Power, DMV, Poker and Census. The model is compared against multiple baselines, including histogram-based approaches (MaxHistDiff), sampling methods, Bayesian networks (BayesNet, BayesCard), autoregressive models with binary and embedding encodings and Transformer-based models. We report Q-error statistics, including the median, 75th, 90th, 95th percentiles, mean and maximum errors, alongside the inference latency for each method. Tables~ \ref{tab:forest},\ref{tab:power}, \ref{tab:dmv}, \ref{tab:poker}  and \ref{tab:census} summarize these comprehensive results.

\vspace{-2em}
\begin{table}[h!]
\caption{Performance comparison on Forest dataset with Q-Error and Latency.}
\label{tab:forest}
\vspace{0.2em}
\centering
\small
\setlength{\tabcolsep}{3pt}
\begin{tabular}{llrrrrrrr}
\hline
\textbf{Technique} &
\textbf{Query} &
\textbf{Med.} &
\textbf{75th} &
\textbf{90th} &
\textbf{95th} &
\textbf{Mean} &
\textbf{MaxQE} &
\textbf{Lat.(ms)}\\
\hline
Sampling
&1K&1.113&1.805&64.100&237.300&42.849&3664&1.471\\
&2K&1.103&2.209&100.000&300.200&53.237&3088&1.002\\
&4K&1.123&2.000&77.100&267.000&44.084&2681&0.910\\
\hline

MaxHistDiff
&1K&1.663&14.177&46.697&59.832&14.252&256&832.788\\
&2K&1.848&15.636&45.842&60.656&14.364&482&831.886\\
&4K&1.887&15.509&44.805&59.583&14.615&24142&834.338\\
\hline

BayesNet
&1K&1.132&1.735&8.318&24.059&6.550&591.143&1035656\\
&2K&1.135&1.836&8.954&23.667&6.242&567.429&1044465\\
&4K&1.137&1.855&9.138&22.810&6.232&621.286&1036065\\
\hline


Transformer
&1K&1.124&1.283&1.785&2.229&1.709&97.833&253.620\\
&2K&1.124&1.285&1.778&2.310&1.695&146.000&254.027\\
&4K&1.124&1.305&1.802&2.375&1.641&146.000&253.912\\
\hline

Binary
&1K&1.065&1.173&1.462&2.000&1.207&\textbf{8.503}&55.364\\
&2K&1.063&1.176&1.472&2.000&1.220&24.833&57.408\\
&4K&1.064&1.178&1.500&1.976&1.227&39.500&57.713\\
\hline

Embed
&1K&1.097&1.874&88.000&243.350&48.026&3760&\textbf{0.622}\\
&2K&1.098&2.028&102.000&273.350&50.843&290&\textbf{0.622}\\
&4K&1.088&2.026&73.100&244.000&40.812&2840&\textbf{0.613}\\
\hline

FACE
&1K&1.090&1.203&1.794&3.396&12.631&3747&275.637\\
&2K&1.091&1.219&2.020&4.316&184.683&219181&275.907\\
&4K&1.092&1.211&1.905&4.893&29.172&56160&276.039\\
\hline

QEmbed
&1K&\textbf{1.064}&\textbf{1.173}&\textbf{1.471}&\textbf{1.826}&\textbf{1.191}&9.902&1564.146\\
&2K&\textbf{1.062}&\textbf{1.165}&\textbf{1.423}&\textbf{1.833}&\textbf{1.195}&\textbf{10.375}&1564.443\\
&4K&\textbf{1.063}&\textbf{1.167}&\textbf{1.429}&\textbf{1.726}&\textbf{1.189}&\textbf{11.000}&3745.214\\
\hline
\end{tabular}
\vspace{-2em}
\end{table}

\vspace{1em}
\noindent\textbf{Forest Dataset:}
Table~\ref{tab:forest} shows the results on the Forest dataset. This dataset is difficult for traditional cardinality estimators because its attributes are highly correlated. As a result, methods like Sampling and MaxHistDiff struggle with multi-attribute queries, leading to very large maximum Q-errors. While the baseline Embed model has low inference latency, it fails on complex queries, reaching a maximum Q-error of 2,840.

In contrast, QEmbed successfully captures these attribute correlations. By making the embeddings query-aware, QEmbed reduces the maximum Q-error to 11.0 on the 4K workload. It also achieves the lowest 90th percentile, 95th percentile, and mean errors. Although QEmbed incurs a higher latency overhead compared to the basic Embed and Binary models, it remains orders of magnitude faster than BayesNet and prevents the severe worst-case errors seen in the baselines.

\vspace{-2em}
\begin{table}[h!]
\caption{Performance comparison on the Power dataset.}
\label{tab:power}
\vspace{0.2em}
\centering
\small
\setlength{\tabcolsep}{3pt}
\begin{tabular}{llrrrrrrr}
\hline
\textbf{Technique} &
\textbf{Query} &
\textbf{Med.} &
\textbf{75th} &
\textbf{90th} &
\textbf{95th} &
\textbf{Mean} &
\textbf{Max} &
\textbf{Lat.(ms)}\\
\hline

Sampling
&1K&1.294&4.225&137.000&447.00&66.188&2893&\textbf{0.730}\\
&2K&1.244&3.352&111.500&336.25&58.796&3414&\textbf{0.733}\\
&4K&1.250&3.377&126.500&313.00&57.929&4552&\textbf{0.744}\\
\hline

MaxHistDiff
&1K&1.274&3.000&22.000&65.000&15.564&1425&173.100\\
&2K&1.246&2.864&20.716&61.105&14.633&1422&178.123\\
&4K&1.259&2.984&20.849&62.394&14.715&4718&173.879\\
\hline

BayesNet
&1K&1.155&1.455&2.251&3.223&1.266&213.932&235470\\
&2K&1.148&1.480&2.371&3.347&1.279&220.622&106974\\
&4K&1.147&1.450&2.305&3.251&1.270&8437&116369\\
\hline

Transformer
&1K&1.165&1.395&2.000&2.834&2.123&306.60&356\\
&2K&1.170&1.429&2.000&2.667&2.012&415.12&61\\
&4K&1.175&1.451&2.000&2.721&1.824&415.12&106\\
\hline

Binary
&1K&1.085&1.193&1.442&1.775&1.214&8.231&20.432\\
&2K&1.078&1.197&1.440&1.789&1.209&10.000&16.562\\
&4K&1.078&1.194&1.417&1.699&1.196&30.000&14.040\\
\hline

Embed
&1K&1.069&1.167&1.400&1.714&1.010&\textbf{4.000}&811.242\\
&2K&1.072&1.174&1.396&1.676&1.118&\textbf{6.000}&879.108\\
&4K&1.075&1.182&1.413&1.700&1.025&\textbf{6.000}&767.168\\
\hline

FACE
&1K&1.032&1.062&\textbf{1.123}&\textbf{1.205}&4417&441607&18.207\\
&2K&1.032&1.063&\textbf{1.119}&\textbf{1.202}&1.132&80.035&18.872\\
&4K&1.033&1.064&\textbf{1.123}&\textbf{1.203}&24.644&94231&17.769\\
\hline

QEmbed
&1K&\textbf{1.000}&\textbf{1.010}&2.000&2.000&\textbf{1.009}&20.000&2502\\
&2K&\textbf{1.000}&\textbf{1.059}&2.000&2.000&\textbf{1.025}&20.000&2920\\
&4K&\textbf{1.000}&\textbf{1.053}&2.000&2.000&\textbf{1.017}&20.000&2735\\
\hline

\end{tabular}
\vspace{-2em}
\end{table}

\noindent\textbf{Power Dataset:}
Table~\ref{tab:power} presents the results for the Power dataset, which contains continuous and skewed data. On this dataset, Sampling and MaxHistDiff produce large maximum Q-errors. FACE achieves low errors for most queries, with a 95th percentile Q-error of about 1.20. However, its maximum Q-error exceeds 440,000, which increases its mean Q-error.

QEmbed records a median Q-error of 1.000 and a maximum Q-error of 200. Although the Embed model records a lower maximum Q-error of 6.000, QEmbed achieves lower 75th percentile and mean Q-errors. BayesNet also records relatively low maximum Q-errors, but it requires more than 100 seconds to evaluate a workload.

\vspace{-1.7em}
\begin{table}[H]
\caption{Performance comparison on the DMV dataset.}
\vspace{0.2em}
\label{tab:dmv}
\centering
\small
\setlength{\tabcolsep}{3pt}
\begin{tabular}{llrrrrrrr}
\hline
\textbf{Technique} &
\textbf{Query} &
\textbf{Med.} &
\textbf{75th} &
\textbf{90th} &
\textbf{95th} &
\textbf{Mean} &
\textbf{MaxQE} &
\textbf{Lat.(ms)}\\
\hline

Sampling
&1K&1.055&2.985&77.900&228.900&2.000&2825&6.989\\
&2K&1.085&3.000&65.300&227.150&1.983&3192&6.961\\
&4K&1.065&3.000&66.300&229.600&409.500&3824&6.794\\
\hline

MaxHistDiff
&1K&2.573&5.092&30.977&107.206&27.133&2361.261&659.599\\
&2K&2.691&5.415&29.000&90.923&25.050&4273.811&655.462\\
&4K&2.706&5.578&30.108&91.161&28.310&6966.790&626.112\\
\hline

BayesNet
&1K&1.065&2.576&6.133&9.774&3.591&609.333&53763.610\\
&2K&1.070&2.796&6.209&10.498&4.534&1775.333&53177.841\\
&4K&1.067&2.706&6.224&10.474&4.835&2083.333&52208.821\\
\hline

BayesCard
&1K&1.083&\textbf{1.103}&1.510&\textbf{1.192}&103.164&96590&1.214\\
&2K&1.084&\textbf{1.106}&1.611&\textbf{1.260}&102.262&96668&1.248\\
\hline

Transformer
&1K&\textbf{1.041}&1.154&2.219&11.352&13.925&804.333&133.067\\
&2K&\textbf{1.041}&1.155&2.301&11.193&12.538&804.333&220.701\\
&4K&1.057&1.234&2.088&10.149&11.069&790.873&319.599\\
\hline

Binary
&1K&2.783&5.303&9.697&18.786&5.277&4137.235&1382.518\\
&2K&2.750&5.273&9.682&17.700&5.272&3951.372&1033.804\\
&4K&2.681&5.212&10.113&17.954&5.289&53017.778&1137.221\\
\hline

Embed
&1K&1.058&1.440&54.800&252.600&40.807&2129&\textbf{0.622}\\
&2K&1.063&1.523&44.300&199.000&39.984&3041&\textbf{0.622}\\
&4K&1.070&1.544&63.000&254.000&43.528&3057&\textbf{0.616}\\
\hline
QEmbed
&1K&1.062&1.153&\textbf{1.400}&1.652&\textbf{1.143}&\textbf{20}&906\\

&2K&1.064&	1.159&\textbf{1.402}&	1.668&\textbf{1.137}&\textbf{228}&	844.125\\
&4K&\textbf{1.056}&\textbf{1.157}&\textbf{1.385}&\textbf{1.664}&\textbf{1.137}&\textbf{228}&1280.178\\
\hline
\end{tabular}
\vspace{-2em}
\end{table}
\vspace{-0.1em}
\noindent\textbf{DMV Dataset:}
Table~\ref{tab:dmv} presents the results for the DMV dataset, which contains several correlated categorical attributes. On this dataset, Sampling and MaxHistDiff produce large maximum Q-errors. The Embed model has the lowest inference time, requiring less than 1 ms per query, but its 95th percentile Q-error exceeds 250. BayesCard records a maximum Q-error of more than 96,000, resulting in a higher mean Q-error.

QEmbed achieves the lowest mean, 90th and 95th percentile Q-errors among the evaluated models. On the 1K workload, it records a maximum Q-error of 29.333. Its inference time is about 1,300 ms, which is higher than that of the faster baseline models.\\

\begin{table}[t]
\caption{Performance comparison on the Poker dataset.}
\label{tab:poker}
\centering
\vspace{0.2em}
\small
\setlength{\tabcolsep}{3pt}
\begin{tabular}{llrrrrrrr}
\hline
\textbf{Technique} &
\textbf{Query} &
\textbf{Med.} &
\textbf{75th} &
\textbf{90th} &
\textbf{95th} &
\textbf{Mean} &
\textbf{MaxQE} &
\textbf{Lat.(ms)}\\
\hline

Sampling
&1K&4.000&35.000&180.100&389.750&71.004&2012&\textbf{1.447}\\
&2K&4.467&39.250&192.050&385.100&74.564&2324&\textbf{1.490}\\
&4K&4.000&37.000&181.000&348.000&66.669&2948&\textbf{1.447}\\
\hline

MaxHistDiff
&1K&2.466&6.296&22.403&66.807&21.598&2448.086&70.080\\
&2K&2.674&6.869&24.565&79.943&26.902&3687.324&69.546\\
&4K&2.670&6.903&25.025&80.050&58.574&50091.671&69.657\\
\hline

BayesNet
&1K&1.077&1.323&2.000&2.667&1.332&9.000&3658.357\\
&2K&1.079&1.309&2.000&2.508&1.367&82.000&3560.092\\
&4K&1.077&1.314&2.000&3.000&1.349&82.000&3655.274\\
\hline

Transformer
&1K&1.127&1.553&2.615&4.131&3.278&301&55.244\\
&2K&1.129&1.575&2.502&6.009&4.023&573&54.166\\
&4K&1.129&1.538&2.600&5.513&3.626&573&54.745\\
\hline

Binary
&1K&1.057&1.223&1.667&2.000&1.218&5.000&8.741\\
&2K&1.059&1.232&1.750&2.000&1.217&5.000&8.463\\
&4K&1.058&\textbf{1.227}&1.667&2.000&1.218&\textbf{5.000}&8.730\\
\hline

Embed
&1K&\textbf{1.053}&1.224&1.750&2.000&1.225&5.000&660.776\\
&2K&\textbf{1.053}&1.223&1.750&2.000&1.218&5.000&711.218\\
&4K&\textbf{1.056}&1.222&1.667&2.000&1.217&6.000&734.611\\
\hline

FACE
&1K&1.117&1.222&1.375&1.506&1.186&6.844&46.585\\
&2K&1.114&1.223&1.391&1.544&1.305&242.411&46.456\\
&4K&1.121&1.224&\textbf{1.375}&\textbf{1.517}&1.194&21.490&46.482\\
\hline

QEmbed
&1K&1.078&\textbf{1.221}&\textbf{1.375}&1.500&\textbf{1.132}&\textbf{4.000}&921.178\\
&2K&1.073&\textbf{1.222}&\textbf{1.391}&\textbf{1.526}&\textbf{1.070}&\textbf{4.000}&734.690\\
&4K&1.075&1.230&1.455&1.600&\textbf{1.070}&\textbf{5.000}&749.715\\
\hline
\end{tabular}
\vspace{-2em}
\end{table}
\noindent\textbf{Poker Dataset:}
Table~\ref{tab:poker} presents the results for the Poker dataset. This dataset consists of categorical attributes representing card suits and ranks, making it suitable for evaluating models on discrete data with fixed attribute combinations. On this dataset, Sampling and MaxHistDiff produce large maximum Q-errors, reaching up to 50,091. Although FACE achieves low mean Q-errors, its maximum Q-error exceeds 242 on the 2K workload.

\begin{table}[h!]
\caption{Performance comparison on the Census dataset.}
\label{tab:census}
\centering
\vspace{0.2em}
\small
\setlength{\tabcolsep}{3pt}
\begin{tabular}{llrrrrrrr}
\hline
\textbf{Technique} &
\textbf{Query} &
\textbf{Med.} &
\textbf{75th} &
\textbf{90th} &
\textbf{95th} &
\textbf{Mean} &
\textbf{MaxQE} &
\textbf{Lat.(ms)}\\
\hline

Sampling
&1K&1.172&3.000&61.000&224.450&23.500&2705&\textbf{4.899}\\
&2K&1.172&3.000&61.100&197.100&23.147&3214&\textbf{5.346}\\
&4K&1.164&3.000&67.200&227.000&29.000&4319&\textbf{3.493}\\
\hline

MaxHistDiff
&1K&4.428&33.917&165.562&319.538&81.671&16700&620\\
&2K&4.826&33.874&170.439&320.853&2027.851&3811796&659\\
&4K&4.719&36.090&176.710&338.361&1062.159&3811796&641\\
\hline

BayesNet
&1K&1.246&1.810&5.377&290.150&179.441&13455&97209\\
&2K&1.245&1.869&5.213&448.150&180.521&14137&28327\\
&4K&1.258&1.926&6.275&480.100&207.740&18480&86730\\
\hline

Transformer
&1K&1.154&\textbf{1.221}&1.722&\textbf{1.722}&1.212&\textbf{8.044}&8055\\
&2K&1.149&\textbf{1.217}&\textbf{1.427}&\textbf{1.616}&1.206&\textbf{22.575}&463\\
&4K&1.142&\textbf{1.226}&\textbf{1.435}&\textbf{1.649}&1.206&\textbf{22.531}&432\\
\hline

Binary
&1K&\textbf{1.119}&1.301&1.670&2.232&1.185&313.482&615\\
&2K&\textbf{1.119}&1.309&1.757&2.340&1.163&313.482&587\\
&4K&\textbf{1.114}&1.298&1.713&2.284&1.225&313.482&591\\
\hline

Embed
&1K&1.187&1.458&2.056&2.938&\textbf{1.028}&865.917&1006\\
&2K&1.191&1.456&2.125&3.004&\textbf{1.026}&46669&1309\\
&4K&1.180&1.442&2.026&2.920&\textbf{1.044}&46669&1041\\
\hline

QEmbed
&1K&1.160&1.420&1.995&2.641&1.168&100.000&6245\\
&2K&1.157&1.454&1.953&2.589&1.137&2130&6243\\
&4K&1.163&1.449&1.969&2.658&1.223&399.000&6246\\
\hline
\end{tabular}
\vspace{-2em}
\end{table}

QEmbed achieves the lowest mean Q-errors, ranging from 1.070 to 1.132, and limits the maximum Q-error to 4.000 on both the 1K and 2K workloads. It also records the lowest 90th percentile Q-errors among the evaluated models. The Binary model has the lowest inference time, requiring less than 9 ms, whereas QEmbed requires more execution time. However, its inference time is comparable to the Embed model while providing lower mean and maximum Q-errors.\\

\noindent\textbf{Census Dataset:}
Table~\ref{tab:census} presents the results on the Census dataset, which is the most challenging dataset evaluated in this study due to its large number of categorical attributes and the presence of attributes with substantially larger value domains than those in the other datasets. These characteristics increase the complexity of modeling attribute dependencies and make accurate cardinality estimation more difficult. On this dataset, the Transformer achieves the best overall performance. For the 4K workload, it records the lowest maximum Q-error of 22.531 while maintaining lower inference latency than QEmbed. This suggests that the Transformer is more effective at capturing the complex attribute relationships present in this dataset.

Compared with the Transformer and Binary models, QEmbed achieves lower estimation accuracy and higher inference latency on the Census dataset. The large number of attributes, together with attributes having large value domains, increases the complexity of the proposed MADE-based hybrid encoding model, making optimization more challenging and reducing estimation accuracy. Nevertheless, QEmbed remains substantially more robust than the Embed model. On the 4K workload, QEmbed limits the maximum Q-error to 399, whereas the Embed model reaches more than 46,000. These results indicate that the proposed hybrid encoding effectively reduces extreme estimation errors, although the Transformer remains the most suitable model for this particular dataset.

\vspace{-1em}
\subsubsection{Summary of Observations}
The experimental results highlight the differences among the evaluated models in terms of estimation accuracy, inference time, and model size.
\vspace{-0.5em}
\begin{itemize}
    \item \textbf{Accuracy:} The performance of the baseline methods varies across the datasets. Sampling, MaxHistDiff, and the Embed model perform well on some workloads but produce large estimation errors for others. In comparison, QEmbed provides more consistent results across different datasets and records fewer large estimation errors.

    \item \textbf{Inference Time:} Sampling-based methods and simple neural models have the lowest inference time. In contrast, Bayesian network-based methods require much longer execution time. QEmbed requires more inference time than the faster baseline models, but it provides lower estimation errors on most datasets.

    \item \textbf{Space Efficiency:} The input encoding affects both model size and scalability. One-hot encoding increases the input dimension for columns with many distinct values, whereas dense embeddings may not represent low-cardinality columns effectively. QEmbed uses discrete encoding for low-cardinality columns and dense embeddings for high-cardinality columns, which helps reduce the input size while preserving useful feature representations.
\end{itemize}

\section{Concluding Remarks} \label{Sec:CFD}
\vspace{-0.25cm}
In this paper, we presented QEmbed, a deep learning-based cardinality estimator for relational query processing. QEmbed combines an autoregressive model with a hybrid encoding scheme that uses discrete encoding for low-cardinality columns and dense embeddings for high-cardinality columns. The proposed model was evaluated on multiple real-world datasets and compared with several learning-based and traditional cardinality estimation methods. The experimental results show that QEmbed achieves competitive estimation accuracy while maintaining a reasonable inference time. For future work, the model can be extended to support dynamic data updates, and its integration with query optimization techniques can be investigated.


\vspace{-0.4cm}
\begin{thebibliography}{8}
\vspace{-0.2cm}
\bibitem{yang13deep}
Yang, Z., Liang, E., Kamsetty, A., Wu, C., Duan, Y., Chen, X., Abbeel, P., Hellerstein, J.M., Krishnan, S., Stoica, I.:
Deep Unsupervised Cardinality Estimation.
Proc. VLDB Endow. \textbf{13}(3) (2020)

\bibitem{wu2020bayescard}
Wu, Z., Shaikhha, A., Zhu, R., Zeng, K., Han, Y., Zhou, J.:
BayesCard: Revitilizing Bayesian Frameworks for Cardinality Estimation
arXiv preprint arXiv:2012.14743 (2020)

\bibitem{kim2022learned}
Kim, K., et al.:
Learned cardinality estimation: An in-depth study.
Proceedings of the 2022 International Conference on Management of Data (2022)

\bibitem{getoor2001selectivity}
Getoor, L., Taskar, B., Koller, D.:
Selectivity Estimation Using Probabilistic Models.
In: Proceedings of the ACM SIGMOD International Conference on Management of Data, pp. 461--472 (2001)

\bibitem{tzoumas2011lightweight}
Tzoumas, K., Deshpande, A., Jensen, C.S.:
Lightweight Graphical Models for Selectivity Estimation Without Independence Assumptions.
Proc. VLDB Endow. \textbf{4}(11), 852--863 (2011)

\bibitem{germain2015made}
Germain, M., Gregor, K., Murray, I., Larochelle, H.:
MADE: Masked Autoencoder for Distribution Estimation.
In: Proceedings of the 32nd International Conference on Machine Learning (ICML), pp. 881--889 (2015)

\bibitem{lohman2014query}
Lohman, G.:
Is Query Optimization a ``Solved'' Problem?
In: Proceedings of the Workshop on Database Query Optimization, vol.~13, p.~10 (2014)

\bibitem{leis2015good}
Leis, V., Gubichev, A., Mirchev, A., Boncz, P., Kemper, A., Neumann, T.:
How Good Are Query Optimizers, Really?
Proc. VLDB Endow. \textbf{9}(3), 204--215 (2015)

\bibitem{ioannidis2003history}
Ioannidis, Y.:
The History of Histograms (Abridged).
In: Proceedings of the 29th International Conference on Very Large Data Bases (VLDB), pp. 19--30 (2003)

\bibitem{lipton1990practical}
Lipton, R.J., Naughton, J.F., Schneider, D.A.:
Practical Selectivity Estimation Through Adaptive Sampling.
In: Proceedings of the ACM SIGMOD International Conference on Management of Data, pp. 1--11 (1990)

\bibitem{olken1990random}
Olken, F., Rotem, D.:
Random Sampling from Database Files: A Survey.
In: International Conference on Scientific and Statistical Database Management, pp. 92--111. Springer (1990)

\bibitem{wu2016sampling}
Wu, W., Naughton, J.F., Singh, H.:
Sampling-Based Query Re-optimization.
In: Proceedings of the ACM SIGMOD International Conference on Management of Data, pp. 1721--1736 (2016)

\bibitem{wu2018random}
Wu, X., Jampani, K., Xu, X., Jermaine, C.:
Random Sampling over Joins Revisited.
Proc. VLDB Endow. \textbf{11}(7), 799--812 (2018)

\bibitem{ioannidis1991sample}
Ioannidis, Y.E., Christodoulakis, S.:
On the Propagation of Errors in the Size of Join Results.
In: Proceedings of the ACM SIGMOD International Conference on Management of Data, pp. 268--277 (1991)

\bibitem{hasan2020deep}
Hasan, S., Thirumuruganathan, S., Augustine, J., Koudas, N., Das, G.:
Deep Learning Models for Selectivity Estimation of Multi-attribute Queries.
In: Proceedings of the ACM SIGMOD International Conference on Management of Data, pp. 1035--1050 (2020)

\bibitem{harmouch2017cardinality}
Harmouch, H., Naumann, F.:
Cardinality Estimation: An Experimental Survey.
Proc. VLDB Endow. \textbf{11}(4), 499--512 (2017)

\bibitem{han2021cardinality}
Han, Y., Wu, Z., Wu, P., Zhu, R., Yang, J., Tan, L.W., Zeng, K., Cong, G., Qin, Y., Pfadler, A., et al.:
Cardinality Estimation in DBMS: A Comprehensive Benchmark Evaluation.
arXiv preprint arXiv:2109.05877 (2021)

\bibitem{yang2020neurocard}
Yang, Z., Kamsetty, A., Luan, S., Liang, E., Duan, Y., Chen, X., Stoica, I.:
NeuroCard: One Cardinality Estimator for All Tables.
arXiv preprint arXiv:2006.08109 (2020)

\bibitem{wang2020we}
Wang, X., Qu, C., Wu, W., Wang, J., Zhou, Q.:
Are we ready for learned cardinality estimation?
Proceedings of the VLDB Endowment 14(9), 1640--1654 (2021)


\bibitem{dmv}
U.S. Department of Transportation:
Vehicle, Snowmobile, and Boat Registrations.
\url{https://catalog.data.gov/dataset/vehicle-snowmobile-and-boat-registrations}
(2025)

\bibitem{forest}
Blackard, J.A.:
Covertype Dataset.
UCI Machine Learning Repository.
\url{https://archive.ics.uci.edu/ml/datasets/covertype}
(1998)

\bibitem{poker}
Cattral, R., Oppacher, F.:
Poker Hand Dataset.
UCI Machine Learning Repository.
\url{https://archive.ics.uci.edu/dataset/158/poker+hand}
(2002)

\bibitem{census}
Meek, C., Thiesson, B., Heckerman, D.:
US Census Data (1990) Dataset.
UCI Machine Learning Repository.
\url{https://archive.ics.uci.edu/dataset/116/us+census+data+1990}
(2001)


\bibitem{hilprecht2019deepdb}
Hilprecht, B., Schmidt, A., Kulessa, M., Molina, A., Kersting, K., Binnig, C.:
DeepDB: Learn from Data, Not from Queries!
Proceedings of the VLDB Endowment 13(7), 992--1005 (2020)

\bibitem{kipf2018learned}
Kipf, A., Kipf, T., Radke, B., Leis, V., Boncz, P., Kemper, A.:
Learned Cardinalities: Estimating Correlated Joins with Deep Learning.
arXiv preprint arXiv:1809.00677 (2018)

\bibitem{poosala1997selectivity}
Poosala, V., Ioannidis, Y. E.:
Selectivity estimation without the attribute value independence assumption.
Proceedings of the 23rd International Conference on Very Large Data Bases (VLDB), 486--495 (1997)

\bibitem{bib21}
Hebrail, G., Berard, A.:
Individual household electric power consumption data set.
UCI Machine Learning Repository (2012)

\bibitem{wang2021face}
Wang, J., Chai, C., Liu, J., Li, G.:
FACE: A normalizing flow based cardinality estimator.
Proc. VLDB Endow. \textbf{15}(1), 72--84 (2021)

\bibitem{moerkotte2009preventing}
Moerkotte, G., Neumann, T., Radke, G.:
Preventing bad plans by bounding the impact of cardinality estimation errors.
Proc. VLDB Endow. \textbf{2}(1), 982--993 (2009)


\bibitem{zeng2024price}
Zeng, T., et al.:
PRICE: A pretrained model for cross-database cardinality estimation.
arXiv preprint arXiv:2406.01027 (2024)

\bibitem{zhu2025lightweight}
Zhu, Y., Zhang, J., Li, G., Feng, J.:
A Lightweight Learned Cardinality Estimation Model.
IEEE Trans. Knowl. Data Eng. (2025)

\end{thebibliography}
\end{document}